\documentclass[a4paper]{jpconf}
\usepackage{graphicx}
\usepackage{amscd}
\usepackage{amsmath}
\usepackage{amssymb}

\newcommand{\g}{{ \mathfrak g}}
\newcommand{\beq}{\begin{equation}}
\newcommand{\eeq}{\end{equation}}
\newcommand{\beqa}{\begin{eqnarray}}
\newcommand{\eeqa}{\end{eqnarray}}
\newcommand{\noi}{\noindent}
\newcommand{\nn}{\nonumber}

\begin{document}
\title{Hidden quartic symmetry in $N=2$ supersymmetry}

\author{Rutwig Campoamor-Stursberg\dag}
\address{\dag\ I.M.I-U.C.M,
Plaza de Ciencias 3, E-28040 Madrid, Spain}
\ead{rutwig@pdi.ucm.es}

\author{Michel Rausch de Traubenberg\ddag}
\address{\ddag IPHC-DRS, UdS, CNRS, IN2P3, 23  rue du Loess,
F-67037 Strasbourg Cedex, France}
\ead{Michel.Rausch@IReS.in2p3.fr}

\begin{abstract}
It is shown that for $N=2$ supersymmetry a hidden symmetry arises
from the hybrid structure of a quartic algebra. The implications
for invariant Lagrangians and multiplets are explored.
\end{abstract}

\section{Introduction}

The consideration of quadratic algebras $-$ Lie (super-)algebras
$-$ largely dominates the algebraic formalism underlying
theoretical physics, although non-quadratic structures emerge, in
more or less natural form, in various descriptions of physical
phenomena. For instance, generalized algebraic approaches are
given by the $n$-linear algebras in Quantum Mechanics \cite{n},
ternary structures in the description of multiple $M_2-$branes
\cite{gbl,az} or higher order extensions of the Poincar\'e algebra
\cite{flie}.

The notion of  Lie algebras of order $F>2$  was first considered
and later intensively studied in \cite{flie,gr,hopf,color},
motivated by the fact that these structures, beyond their formal
mathematical properties, constitute a cornerstone for the
construction of higher order extensions of the Poincar\'e algebra.
In this context, a  specific cubic extension in arbitrary
space-time dimension was shown to be of interest in the frame of
Quantum Field Theory \cite{cubic1,cubic2,p-form}. Finally, it was
realised that it was possible to associate a group \cite{hopf} and
an adapted superspace associated to these structures \cite{super}.

Probably, the main technical difficulty related to Lie algebras of
order $F$ is their hybrid structure: the ``algebra'' is partially
quadratic and partially of order $F$. In this work, we would like
to show that to graded Lie superalgebras of certain form, one can
naturally associate a quartic algebra. It is  shown that along the
lines of this construction, one can associate to $N=2$
supersymmetry a quartic extension of the Poincar\'e algebra. This
construction indicates that some kind of hidden quartic symmetry
appears in usual supersymmetry, which further means that invariant
Lagrangians constructed so far are also invariant under the
induced quartic structure.

\noindent We illustrate the fact that on the top of the
representations of supersymmetry, a hierarchy of representations
can be constructed. The work presented in this note was obtained
in \cite{quart} in more detail.

\section{Lie algebras of order four $-$ quartic extensions of the Poincar\'e algebra}

There are various types of extensions of Lie algebras that can be
considered. The case under inspection here, enabling us to
construct non-trivial extensions of the Poincar\'e algebra, are
related to the quartic case, for which we recall the main
properties properties.

\noindent The vector space $\g=\g_0 \oplus \g_1 $ with basis $
\big\{ X_i, i=1,\cdots, \dim \g_0\big\}, \big\{ Y_a, a=1,\cdots,
\dim \g_1\big\} $ is called an elementary Lie algebra of order
four if it satisfies the following brackets \cite{flie}

\beqa \label{eq:4-alg} \big[X_i,X_j\big]=f_{ij}{}^k X_k, & & \ \
\big[X_i,Y_a\big]=R_{ia}{}^b Y_b, \ \ \nonumber \\
\big\{Y_{a_i},Y_{a_2},Y_{a_3},Y_{a_4}\big\}&=& \sum
\limits_{\sigma \in S_4}
Y_{\sigma(a_1)}Y_{\sigma(a_2)}Y_{\sigma(a_3)}Y_{\sigma(a_4)} =
Q_{a_1 a_2 a_3 a_4}{}^i X_i, \eeqa

\noi $S_4$ being the permutation group with four elements.
In
addition, we have also the following generalised Jacobi
identities:

\beqa \label{eq:J}
 \big[Y_{a_1},
\big\{Y_{a_2},Y_{a_3},Y_{a_4},Y_{a_5} \big\}\big]+ \big[Y_{a_2},
\big\{Y_{a_3},Y_{a_4},Y_{a_5},Y_{a_1} \big\}\big]+ \big[Y_{a_3},
\big\{Y_{a_4},Y_{a_5},Y_{a_1},Y_{a_2} \big\}\big]+\nonumber \\
\big[Y_{a_4}, \big\{Y_{a_5},Y_{a_1},Y_{a_2},Y_{a_3} \big\}\big] +
\big[Y_{a_5}, \big\{Y_{a_1},Y_{a_2},Y_{a_3},Y_{a_4} \big\}\big]
=0. \eeqa

Let us note that the structure defined by equations
\eqref{eq:4-alg} and \eqref{eq:J} is neither an algebra nor a
$4-$algebra in the usual sense, but a kind of hybrid structure.
Some of the brackets will be quadratic $[\g_0, \g_0] \subseteq
\g_0, [\g_0, \g_1] \subseteq \g_1$, while some others will be
quartic $\big\{\g_1,\g_1,\g_1,\g_1 \big\} \subseteq \g_0$. This
feature obviously generates the question whether from this hybrid
structure we can extract some additional properties that cannot be
codified either by the binary or quartic structure alone.

\noindent In the preceding context, the quartic extensions of the
Poincar\'e algebra in $D=4$ dimensions are realised by means of
two Majorana spinors. Using the $\mathfrak{sl}(2,\mathbb C) \cong
\mathfrak{so}(1,3)$ notations of dotted and undotted indices, a
left-handed spinor is given by $\psi_L{}^\alpha$ and  a
right-handed spinor by $\bar \psi_R{}_{\dot \alpha}$.  The spinor
conventions to raise/lower indices are the following:
$\psi_L{}_\alpha =\varepsilon_{\alpha\beta}\psi_L{}^\beta$,
$\psi_L{}^\alpha =\varepsilon^{\alpha\beta}\psi_L{}_\beta$,
$\bar\psi_R{}_{\dot\alpha}=\varepsilon_{\dot\alpha
\dot\beta}\bar\psi_R{}^{\dot\beta}$, $\bar\psi_R{}^{\dot\alpha}
=\varepsilon^{\dot\alpha\dot\beta}\bar\psi_R{}_{\dot\beta}$ with
$(\psi_\alpha)^\star =\bar\psi_{\dot\alpha}$, $\varepsilon_{12} =
\varepsilon_{\dot 1\dot 2}=1$, $\varepsilon^{12} =
\varepsilon^{\dot 1\dot 2}=-1$. The $4D$ Dirac matrices,  in the
Weyl representation,  are
\begin{equation}
\label{eq:gamma} \Gamma^\mu = \left(
\begin{array}{cc}
 0 & \sigma^\mu\\
 \bar\sigma^\mu & 0
\end{array}\right),
\end{equation}
with $\sigma^\mu{}_{\alpha\dot\alpha}=(1,\sigma^i ),
 \; \bar\sigma^{\mu\dot\alpha\alpha}=
 (1,-\sigma^i)$,
$\sigma^i$ ($i=1,2,3$) being the Pauli matrices. With these
notations, we introduce two series of Majorana spinors
$Q^I{}_{\alpha}, \bar Q_I{}_{\dot \alpha}$  satisfying the
relation $(Q^I{}_{\alpha})^\dag= \bar Q_I{}_{\dot \alpha}$. The
Lie algebra of order four with $\g_0=I{\mathfrak{so}}(1,3) $ (the
Poincar\'e algebra) and $\g_1= \big<Q^I{}_\alpha, \bar Q_{I \dot
\alpha}\big>$ define the following quartic extension of the
Poincar\'e algebra (we only give the
quartic brackets explicitly)

\beqa \label{4poin} \fl
\left\{Q^{I_1}{}_{\alpha_1},Q^{I_2}{}_{\alpha_2},
Q^{I_3}{}_{\alpha_3}, Q^{I_4}{}_{\alpha_4}
\right\}&=&0, \nonumber \\
%%%%%
\fl \left\{Q^{I_1}{}_{\alpha_1},Q^{I_2}{}_{\alpha_2},
Q^{I_3}{}_{\alpha_3}, \bar{Q}_{I_4}{}_{\dot \alpha_4} \right\}&=&
2i\Big(\delta^{I_1}{}_{I_4} \varepsilon^{I_2 I_3}
\varepsilon_{\alpha_2 \alpha_3} \sigma^\mu{}_{\alpha_1 \dot
\alpha_4} +\delta^{I_2}{}_{I_4} \varepsilon^{I_1 I_3}
\varepsilon_{\alpha_1 \alpha_3} \sigma^\mu{}_{\alpha_2 \dot \alpha_4}  \\
&+&\delta^{I_3}{}_{I_4} \varepsilon^{I_1 I_2}
\varepsilon_{\alpha_1 \alpha_2} \sigma^\mu{}_{\alpha_3 \dot
\alpha_4}\Big) P_\mu,
\nonumber \\
\fl \left\{Q^{I_1}{}_{\alpha_1},Q^{I_2}{}_{\alpha_2},
\bar{Q}_{I_3}{}_{\dot \alpha_3}, \bar{Q}_{I_4}{}_{\dot \alpha_4}
\right\}&=& 0, \nonumber \eeqa

\noi the remaining brackets involving three $\bar Q$ and one $Q$
or four $\bar Q$ being obtained immediately (the tensor
$\varepsilon^{IJ}$ is defined by
$-\varepsilon^{12}=\varepsilon^{21}=\varepsilon_{12}=-\varepsilon_{21}=1$).

As noted in the previously, the quartic extension of the
Poincar\'e algebra obtained is neither an algebra nor a
four-algebra. This feature represents one of the difficulties to
handle with these algebraic structures. Consequently one natural
question we should address concerns the possibility to associate
appropriate quadratic structures to Lie algebras of order four.
Its has to be mentioned that a similar analysis has been performed
for Lie algebras of order three, were it has been shown that, no
possibility to construct an associated quartic structure exists.

\section{Quartic structures associated to Lie superalgebras}
As mentioned earlier, higher order extensions are no fully
satisfactory, in spite of various interesting results derived for
them \cite{cubic1,cubic2,p-form}). Thus one may wonder whether or
not some quadratic structure should be related to the algebra
\eqref{4poin}. This question is partially motivated by the fact
that for some ternary algebras \cite{fo} of the Filippov type
considered in the Bagger-Lambert-Gustavsson model are equivalent
to certain Lie (super-)algebras \cite{gbl,f,fo}.

\noindent We consider the  $\mathbb Z_2 \times \mathbb Z_2-$graded
Lie superalgebra \beqa\g= \big(\g_{(0,0)} \oplus \g_{(1,1)}\big)
\oplus\big(\g_{(1,0)} \oplus \g_{(0,1)}\big), \eeqa where $(a,b)
\in \mathbb Z_2 \times \mathbb Z_2$ and $\g_{(a,b)}$ is even
(resp. odd) when $a+b=0 \;\rm{mod}\;2$ (resp. $a+b=1 \;\rm{mod} \;
2$).
 Introduce the corresponding bases for the grading
blocks: \beqa \label{eq:gradedLie}
\begin{array}{ll}
\g_{(1,1)}=\langle B_i, i=1,\cdots,\dim \g_{(0,0)}\rangle,&
\g_{(0,0)}=\langle Z\rangle, \\
\g_{(1,0)}=\langle F^+_a, a=1,\cdots,\dim \g_{(1,0)}\rangle,&
\g_{(0,1)}=\langle F^-_a, a=1,\cdots,\dim \g_{(0,1)}\rangle,
\end{array}
\eeqa  the corresponding commutation relations are \beqa
\label{eq:gradedsuper}
\begin{array}{ll}
\big[B_i,B_j\big]=f_{ij}{}^k B_k,&[B_i,Z]=0,\\
\big[B_i,F^{\pm}_a]=R^{\pm}_i{}_a{}^b F^{\pm}_b,& \big[Z,F^{\pm}_a]=0,\\
\big\{F^{+}_i,F^{-}_j\big\}=Q_{ij}{}^a
B_a,& \big\{F^{\pm}_i,F^{\pm}_j\big\}=g^\pm_{ij} Z.
\end{array}
\eeqa  We mention that, the superalgebra defined by
(\ref{eq:gradedsuper}) satisfies also the appropriate Jacobi
identities (that we do not recall here since they will not be
relevant for our purpose).
  It is important to
notice that $\g_{(0,0)}$ commutes with all remaining factors, in
other words, that $Z$ acts like a central charge.

We now show that to the algebra \eqref{eq:gradedsuper} one can naturally and simply
associate a quartic structure which share some similarities with the algebra
\eqref{eq:4-alg}.
Indeed,  using the
obvious relation, \beqa
\big\{A_1,A_2,A_3,A_4\big\}=\big\{\big\{A_1,A_2\big\},\big\{A_3,A_4\big\}
\big\} + \big\{\big\{A_1,A_3\big\},\big\{A_2,A_4\big\} \big\}+
\big\{\big\{A_1,A_4\big\},\big\{A_2,A_3\big\} \big\}, \nonumber
\eeqa
 the relations
\beqa \label{eq:quart} \fl
\big\{F^+_{a_1},F^+_{a_2},F^+_{a_3},F^+_{a_4}\big\} &=&
\Big(g^+_{a_1 a_2} g^+_{a_3 a_4}+
g^+_{a_1 a_3} g^+_{a_2 a_4}+
g^+_{a_1 a_4} g^+_{a_2 a_3}\Big) Z^2
\nonumber \\
\fl \big\{F^+_{a_1},F^+_{a_2},F^+_{a_3},F^-_{a_4}\big\} &=&
2Z(g^+_{a_1 a_2} Q_{a_3 a_4}{}^i + g^+_{a_1 a_3} Q_{a_2 a_4}{}^i +
g^+_{a_3 a_3} Q_{a_1 a_4}{}^i
) B_i  \\
\fl \big\{F^+_{a_1},F^+_{a_2},F^-_{a_3},F^-_{a_4}\big\}&=&
\Big(Q_{a_1 a_3}{}^i Q_{a_2 a_4}{}^j +
Q_{a_1 a_4}{}^i Q_{a_2 a_3}{}^j \big\{B_i,B_j\big\} +
g^+_{i_1 i_2} g^-_{i_3i_4} Z^2   \
,\nonumber \eeqa (plus similar relations involving either three
$F^-$ and one $F^+$ or four $F^-$) follow at once.

\medskip
Since we are constructing an analogue of the four-Lie algebra
(\ref{eq:4-alg}), we also assume that the algebra
 is partially quadratic and partially quartic. This means that in addition
to the brackets \eqref{eq:quart}, we have also to define the
quadratic brackets $[\g_0,\g_0] \subseteq \g_0, [\g_0,\g_1]
\subseteq \g_1$. We simply assume that these brackets are the same
of the corresponding brackets of the Lie superalgebra. As the
quartic brackets are concerned, we observe that
$\{\g_1,\g_1,\g_1,\g_1\}$ close quadratically in $\g_0$. The next
step in the construction is to impose the Jacobi identities
\eqref{eq:J}. This is an extra condition. Indeed, one can show
that the Jacobi identity of the Lie superalgebras do not reproduce
the generalised Jacobi identity (\ref{eq:J}). This will not lead
to any contradiction since it happens that if we have a finite
dimensional representation of (\ref{eq:quart}), the identities
(\ref{eq:J}) are trivially satisfied. Moreover, for the case under
inspection in this work this will not be a constraint, since the
generalised Jacobi identity will be trivially satisfied as well.
This happens because the four-brackets $\{\g_1,\g_1,\g_1,\g_1\}$
close upon $P_\mu$ or $Z$ (see below) thus we automatically have
$[\{\g_1,\g_1,\g_1,\g_1\},\g_1]=0$.

Finally since  {\it by construction} the relations
\eqref{eq:quart} are just a consequence of \eqref{eq:gradedsuper},
this means that the Lie algebras of the form
\eqref{eq:gradedsuper} present some hidden quartic symmetry. In
some sense we could say that the algebra \eqref{eq:quart} is the
``square'' of the algebra \eqref{eq:gradedLie}. Furthermore, the
fact that quadratic relations imply quartic relations   means that
any representation of the Lie superalgebra (\ref{eq:gradedLie})
will also be a (non-faithful) representation of the quartic
algebra (\ref{eq:quart}). Of course, the converse is not
necessarily true.

\noindent We now focus on the relavant case of the $N=2$
supersymmetric extension of the Poincar\'e algebra with central
charge, and show that it is of the form \eqref{eq:quart}. Indeed,
for the even part, we define $\g_{(1,1)}=
I\mathfrak{so}(1,3)=\langle L_{\mu \nu}, P_\mu\rangle$ to be the
Poincar\'e algebra in four-dimensions and $\g_{(0,0)} = \langle Z
\rangle$ the central charge. Although, for the odd part we
introduce two series of Majorana spinors $Q^I{}_{\alpha}, \bar
Q_I{}_{\dot \alpha}, I=1,2$ ($(Q^I{}_{\alpha})^\dag= \bar
Q_I{}_{\dot \alpha}$) such that $\g_{(1,0)}= \langle
Q^I{}_\alpha\rangle$ and $\g_{(0,1)}= \langle \bar Q_{I \dot
\alpha }\rangle$. Since the $N=2$ supersymmetric extension of the
Poincar\'e algebra takes the form \beqa \Big\{Q^I{}_\alpha,\bar
Q_{J \dot \alpha }\Big\}&=& -2i \delta^I{}_J
\sigma^\mu{}_{\alpha \dot \alpha} P_\mu \nn \\
\Big\{Q^I{}_\alpha, Q^I{}_{ \beta }\Big\}&=& 2Z \varepsilon^{IJ} \varepsilon_{\alpha \beta} \\
\Big\{\bar Q_{I \dot \alpha}, \bar Q_{J \dot \beta }\Big\}&=&
-2Z \varepsilon_{IJ} \varepsilon_{\dot \alpha \dot\beta}, \nn
 \eeqa
which is analogous to \eqref{eq:gradedsuper},
the results of the previous section give rise to the four order quartic extension
of the Poincar\'e algebra
\beqa
\big\{Q^{I_1}{}_{\alpha_1}{},Q^{I_2}{}_{\alpha_2}{},Q^{I_3}{}_{\alpha_3}{},Q{I_4}{}_{\alpha_4}\big\}&=&
2Z^2\Big(\varepsilon_{\alpha_1 \alpha_2} \varepsilon_{\alpha_3
\alpha_4} \varepsilon^{I_1 I_2} \varepsilon^{I_3 I_4}\nonumber \\
&& +\varepsilon_{\alpha_1 \alpha_3} \varepsilon_{\alpha_2 \alpha_4}
\varepsilon^{I_1 I_3} \varepsilon^{I_2 I_4} +\varepsilon_{\alpha_1
\alpha_4} \varepsilon_{\alpha_2 \alpha_3} \varepsilon^{I_1 I_4}
\varepsilon^{I_2 I_3}
\Big), \nonumber \\
%%%%
 \left\{Q^{I_1}{}_{\alpha_1},Q^{I_2}{}_{\alpha_2},
Q^{I_3}{}_{\alpha_3}, \bar{Q}_{I_4}{}_{\dot \alpha_4} \right\}&=&
-2i  Z \Big(\delta^{I_1}{}_{I_4} \varepsilon^{I_2 I_3}
\varepsilon_{\alpha_2 \alpha_3} \sigma^\mu{}_{\alpha_1 \dot
\alpha_4} +\delta^{I_2}{}_{I_4} \varepsilon^{I_1 I_3}
\varepsilon_{\alpha_1 \alpha_3} \sigma^\mu{}_{\alpha_2 \dot \alpha_4}  \nonumber \\
&+&\delta^{I_3}{}_{I_4} \varepsilon^{I_1 I_2}
\varepsilon_{\alpha_1 \alpha_2} \sigma^\mu{}_{\alpha_3 \dot
\alpha_4}\Big)P_\mu ,
\nonumber \\
%%%%
\fl \left\{Q^{I_1}{}_{\alpha_1},Q^{I_2}{}_{\alpha_2},
\bar{Q}_{I_3}{}_{\dot \alpha_3}, \bar{Q}_{I_4}{}_{\dot \alpha_4}
\right\}&=&2  \Big(\delta^{I_1}{}_{I_3}
\delta^{I_2}{}_{I_4}\sigma^\mu{}_{\alpha_1 \dot \alpha_3}
\sigma^\nu{}_{\alpha_2 \dot \alpha_4} +\delta^{I_1}{}_{I_4}
\delta^{I_2}{}_{I_3}\sigma^\mu{}_{\alpha_1 \dot
\alpha_4} \sigma^\nu{}_{\alpha_2 \dot \alpha_3} \Big)P_\mu P_\nu \nonumber \\
&& \hskip-.38truecm +\ 2 Z^2 \varepsilon_{\alpha_1 \alpha_2}
\varepsilon_{\dot \alpha_3 \dot \alpha_4} \varepsilon^{I_1I_2}
\varepsilon_{I_3 I_4}. \nonumber \eeqa

A representation of the super-Poincar\'e algebra will
automatically be a representation of the induced quartic algebra,
as shown in the general case. This fact provides us with an
interesting consequence, namely, that the invariant $N=2$
Lagrangians constructed so far are moreover invariant with respect
to the transformations induced by the quartic algebra. Thus, the
corresponding $N=2$ supermultiplet and their associated
transformations laws will automatically be an invariant multiplet
of the corresponding quartic structure with the same
transformation properties. This construction can thus be
interpreted, in some sense, as a possibility to circumvent the
constraints of the Haag-Lopuszanski-Sohnius theorem \cite{hls}.
This analogy should however not be pushed too far: the
construction of quartic algebras executed in this work depends
essentially on the supersymmetric algebra formalism and the
associated constraints.

\section{Representation of quartic extensions of the Poincar\'e
algebra} As we have seen, the ansatz linking algebras of order
four to Lie superalgebras has remarkable consequences concerning
their respective representation theories, in the sense that
superalgebra representations automatically induce representations
of the order four structures. We point outa that the converse of
this statement is not true. Consider for instance massive
representations. The little algebra is generated
 by  $P^0=-im$ and $Q^I{}_\alpha, \bar Q_{I\dot \alpha}$
and  the
four-brackets take the form \beqa
\big\{Q_{\alpha_1}{}^{I_1},Q_{\alpha_2}{}^{I_2},Q_{\alpha_3}{}^{I_3},Q_{\alpha_4}{}^{I_4}\big\}&=&
2Z^2\Big(\varepsilon_{\alpha_1 \alpha_2} \varepsilon_{\alpha_3
\alpha_4} \varepsilon^{I_1 I_2} \varepsilon^{I_3 I_4}\nonumber \\
&& +\varepsilon_{\alpha_1 \alpha_3} \varepsilon_{\alpha_2 \alpha_4}
\varepsilon^{I_1 I_3} \varepsilon^{I_2 I_4} +\varepsilon_{\alpha_1
\alpha_4} \varepsilon_{\alpha_2 \alpha_3} \varepsilon^{I_1 I_4}
\varepsilon^{I_2 I_3}
\Big), \nonumber \\
%%%%
\fl \left\{Q^{I_1}{}_{\alpha_1},Q^{I_2}{}_{\alpha_2},
Q^{I_3}{}_{\alpha_3}, \bar{Q}_{I_4}{}_{\dot \alpha_4} \right\}&=&
2 m Z \Big(\delta^{I_1}{}_{I_4} \varepsilon^{I_2 I_3}
\varepsilon_{\alpha_2 \alpha_3} \sigma^0{}_{\alpha_1 \dot
\alpha_4} +\delta^{I_2}{}_{I_4} \varepsilon^{I_1 I_3}
\varepsilon_{\alpha_1 \alpha_3} \sigma^0{}_{\alpha_2 \dot \alpha_4}  \nonumber \\
&+&\delta^{I_3}{}_{I_4} \varepsilon^{I_1 I_2}
\varepsilon_{\alpha_1 \alpha_2} \sigma^0{}_{\alpha_3 \dot
\alpha_4}\Big) ,
\nonumber \\
%%%%
\fl \left\{Q^{I_1}{}_{\alpha_1},Q^{I_2}{}_{\alpha_2},
\bar{Q}_{I_3}{}_{\dot \alpha_3}, \bar{Q}_{I_4}{}_{\dot \alpha_4}
\right\}&=&2 m^2 \Big(\delta^{I_1}{}_{I_3}
\delta^{I_2}{}_{I_4}\sigma^0{}_{\alpha_1 \dot \alpha_3}
\sigma^0{}_{\alpha_2 \dot \alpha_4} +\delta^{I_1}{}_{I_4}
\delta^{I_2}{}_{I_3}\sigma^0{}_{\alpha_1 \dot
\alpha_4} \sigma^0{}_{\alpha_2 \dot \alpha_3} \Big) \nonumber \\
&& \hskip-.38truecm +\ 2 Z^2 \varepsilon_{\alpha_1 \alpha_2}
\varepsilon_{\dot \alpha_3 \dot \alpha_4} \varepsilon^{I_1I_2}
\varepsilon_{I_3 I_4}. \nonumber \eeqa If we now make the
following substitutions (analogous to the corresponding
substitution for the $N=2$ supersymmetric extension with central
charge): \beqa
\begin{array}{cc}
a^1=Q^1{}_1-\bar{Q}_{2 \dot 2}\ ,&a^3= Q^1{}_1+\bar{Q}_{2 \dot 2}\ , \\
a^2=Q^1{}_2+ \bar{Q}_{2 \dot 1}\ , &a^4=Q^1{}_2- \bar{Q}_{2 \dot
1} \ ,
\end{array}
\eeqa one observes that $a^1,\cdots,a^4,a^\dag_1,\cdots, a^\dag_4$
generate the Clifford algebra of the polynomial \beqa
P^2(x_1,\cdots,x_4,y^1,\cdots,y^4)&=&\Big(2(2m+Z)x_1 y^1 +
2(2m+Z)x_2 y^2 +2(2m-Z)x_3 y^3 \nonumber \\ &&  + 2(2m-Z)x_4 y^4\Big)^2
\nonumber% \\
\eeqa in the sense that \beqa \label{eq:cliff} \Big(x_I a^I + y^I
a^\dag_I\Big)^4 = P^2(x_1,\cdots,x_4,y^1,\cdots,y^4). \eeqa The
representations of the $N=2$ supersymmetric algebra in four
dimensions are obtained from the study of representations of the
Clifford algebra, {\it i.e.} when the $a$'s   satisfy the
quadratic relation, \beqa \label{eq:cliff2} \Big(x_I a^I + y^I
a^\dag_I\Big)^2 = P(x_1,\cdots,x_4,y^1,\cdots,y^4), \eeqa which is
obviously compatible with (\ref{eq:cliff}). On the contrary, one
can construct representations of (\ref{eq:cliff}) such that the
condition (\ref{eq:cliff2}) is not satisfied.\footnote{The algebra
(\ref{eq:cliff}) is called  the Clifford algebra of the polynomial
$P^2$ \cite{cp}.} It can be shown that to any polynomial $f$, a
Clifford algebra ${\cal C}_f$ can be associated to it, and that a
matrix representation can be obtained \cite{line}. Uniqueness is
however lost for degree higher than two, which complicates
considerably the analysis of representations (see, for instance,
\cite{ineq}). Being still an unsolved problem, some structural
results have been already obtained in the general frame
\cite{dimrep}.

In the context that occupies us, this procedure may give new
representations corresponding to interesting quartic extensions of
the Poincar\'e algebra. The hierarchy of representations  on the
top of the standard representations obtained in supersymmetric
theories might be share some similarities with the parafermionic
extension of the Poincar\'e algebra considered in \cite{j}.

\section{Concluding remarks}
We have pointed out the existence of a formal way to associate a
quartic algebra (which closes with fully symmetric quartic
brackets) to a graded Lie superalgebra of a certain type, their
fundamental interest being their application to standard
supersymmetric theories. This specifically alludes to the fact
that any representation of $N=2$ supersymmetric algebras shares a
hidden symmetry arising from the quartic structure. Further, for
massive representations, it turns out that the role of central
charge is essential to the argument.
\smallskip

\noindent Clearly these results can be generalised to other
space-time dimensions \cite{quart}. For example, it turns out that
that the quartic extensions in ten space-time dimensions is
exceptional and related to type $IIA$ supersymmetry. Although we
have focused on the case where $\g_{(0,0)}$ is one dimensional,
there is no reason for restricting to only one central charge. The
straightforward generalisation to a higher number of charges is
however subjected to finding the appropriate algebraic structures
having a physical significance.

\end{document}